\documentclass[aps,twocolumn]{revtex4}
\usepackage{amsmath}
\usepackage{amssymb}
\usepackage{amsfonts}
\usepackage{slashbox}

\usepackage{graphicx}

\usepackage{amsthm}

\newtheorem*{theorem*}     {Theorem}

\newtheorem*{lemma*}       {Lemma}

\newtheorem*{corollary*}   {Corollary}

\theoremstyle{remark}
\newtheorem*{remark*}      {Remark}

\newtheorem*{example*}     {Example}

\DeclareMathOperator{\Ad}{Ad}

\DeclareMathOperator{\SO}{{\it SO}}

\DeclareMathOperator{\SU}{{\it SU}}

\newcommand{\paren}[1]{{\left( #1 \right)}}

\newcommand{\bracket}[1]{{\left[ #1 \right]}}

\renewcommand{\Re}{{\rm Re}}
\renewcommand{\Im}{{\rm Im}}
\newcommand{\diag}{{\mbox{\rm diag}}}

\newcommand{\mtx}[4]{\left[ \begin{array}{cc} 
        {#1} & {#2} \\ {#3} & {#4} \end{array}\right] }

\makeatletter
\def\loweq@align#1#2{\lower.6ex\vbox{\baselineskip\z@skip\lineskip\z@
    \ialign{$\m@th#1\hfil##\hfil$\crcr#2\crcr=\crcr}}}
\def\lowsim@align#1#2{\lower.6ex\vbox{\baselineskip\z@skip\lineskip\z@
    \ialign{$\m@th#1\hfil##\hfil$\crcr#2\crcr\sim\crcr}}}
\def\geqq{\mathrel{\mathpalette\loweq@align >}}
\def\leqq{\mathrel{\mathpalette\loweq@align <}}
\def\grsim{\mathrel{\mathpalette\lowsim@align >}}
\def\lesssim{\mathrel{\mathpalette\lowsim@align <}}
\def\gsim{\mathrel{\mathpalette\lowsim@align >}}
\def\lsim{\mathrel{\mathpalette\lowsim@align <}}
\def\lddots{\mathinner{\mkern1mu
    \raise\p@\hbox{.}\mkern2mu\raise4\p@\hbox{.}\mkern2mu
    \raise7\p@\vbox{\kern7\p@\hbox{.}}\mkern1mu}}
\makeatother
\newcommand{\grless} 
{ {\, \raise-.24em\hbox{$<$} \hspace{-0.8em} \raise.31em\hbox{$>$}\, } }
\newcommand{\lessgr} 
{ {\, \raise-.24em\hbox{$>$} \hspace{-0.8em} \raise.31em\hbox{$<$}\, } }

\newfont{\bg}{cmr10 scaled\magstep4}                    
\newcommand{\bigzerou}{\smash{\lower1.7ex\hbox{\bg 0}}}

\newcommand{\nn}{\nonumber \\ }
\newcommand{\G}{{\SU(2,2)}}
\newcommand{\Goo}{{\SO(4,2)}}
\newcommand{\Go}{{\SO(4,2)_0}}
\newcommand{\g}{{\alg s\alg u(2,2)}}
\newcommand{\go}{{\alg s\alg o(4,2)}}

\newcommand{\disp}{\displaystyle}

\newcommand{\C}{{\mathbb C}}

\newcommand{\R}{{\mathbb R}}

\newcommand{\crl}[1]{[-\infty,\infty]}

\newcommand{\E}{E} %{{\mathbb E}}

\renewcommand{\Ref}[1]{(\ref{#1})}

\newcommand{\wt}{\widetilde}

\newcommand{\wb}{\overline}
\newcommand{\conj}[1]{#1^*}

\newcommand{\ot}{\otimes}
\newcommand{\da}[1]{#1^\dag}

\newcommand{\pa}[1]{\sigma_{#1}}
\newcommand{\la}[1]{\lambda_{#1}}

\newcommand{\z}[1]{z_{#1}}

\newcommand{\sa}{selfadjoint}
\newcommand{\asa}{anti-selfadjoint}
\newcommand{\alg}[1]{{\mathfrak #1}}

\newcommand{\M}{{\cal M}}

\newcommand{\AdS}{{AdS}}

\newcommand{\pad}[1]{{#1}^{\star}}
\newcommand{\ie}{{i.e.}}

\newcommand{\simu}{\sim}
\newcommand{\sime}{\stackrel{\rm \eta}\sim}
\newcommand{\simh}{\stackrel{H}\sim}
\newcommand{\simsc}{\stackrel{\rm sc}\sim}
\newcommand{\co}{cohomogeneity-one}
\newcommand{\conohyphen}{cohomogeneity one}
\newcommand{\Or}{{\cal O}}
\newcommand{\Sigmaa}{U}
\newcommand{\const}{\rm constant}
\newcommand{\HP}{homogeneity-preserving}
\newcommand{\sfc}{{\cal S}}

\begin{document}

%% preprint number
%%\vspace*{-.5cm}
%%\hspace{13cm} \hbox{OCU-PHYS-276/AP-GR-47}

\title{%
Strings in five-dimensional anti-de Sitter space with a symmetry}
\author{Tatsuhiko Koike}
\email{koike@phys.keio.ac.jp}
\affiliation{Department of Physics, Keio University, Yokohama 
  223--8522 Japan}
\author{Hiroshi Kozaki}
\email{kozaki@adm.niit.ac.jp} 
\affiliation{%Department of Mechanical and Control Engineering, 
  Department of Applied Chemistry and Biotechnology, 
  Niigata Institute of Technology, 
  Kashiwazaki, Niigata 945--1195 Japan}
\author{Hideki Ishihara}
\email{ishihara@sci.osaka-cu.ac.jp}
\affiliation{Department of Mathematics and Physics, 
  Graduate School of Science, Osaka City University, 
  Osaka 558--8585 Japan
}

\date{February, 2008}

\begin{abstract}
  The equation of motion of an extended object in spacetime 
  reduces 
  to an ordinary differential equation 
  in the presence of symmetry. 
  %%which may allow one to obtain exact solutions. 
  By properly defining of the symmetry with 
  notion of cohomogeneity, 
  we discuss the method for classifying all these extended objects. 
  %%We then give a method for solving the trajectory of Nambu-Goto strings
  We carry out the classification for the strings 
  in the five-dimensional anti-de Sitter space by 
  the effective use of the local isomorphism between $\SO(4,2)$ and
  $\SU(2,2)$. 
  In the case where the string is described by the
  Nambu-Goto action, 
  we present a general method for solving the trajectory. 
  We then apply the method to one of the classification cases, 
  where the spacetime naturally obtains a Hopf-like bundle structure, 
  and find a solution. 
  The geometry of the solution is analized and found to be a
  timelike helicoid-like surface. 
\end{abstract}

\maketitle

\section{Introduction}
Existence and dynamics of extended objects 
play important roles in various
stages in cosmology. 
%%Among them 
Examples of extended objects 
include topological defects, 
such as  strings and membranes, 
and the Universe as a whole 
embedded in a higher-dimensional spacetime 
in the context of the brane-world universe model~\cite{RanSun99PRL}. 

The trajectory of an extended object forms a hypersurface 
in the spacetime which is determined by a 
partial differential equation (PDE). 
For example, a test string 
is described by the Nambu-Goto equation which is a PDE in two
dimensions.  
Because the dynamics is more complicated than that of 
a particle, one 
usually cannot obtain general solutions. 
One way to find exact solutions is to assume symmetry. 
The simplest solutions to such a PDE are homogeneous ones, 
in which case the problem reduces to a set of algebraic
equations.  
However, the solutions do not have much variety 
and the dynamics is trivial. 
%%In the example of Nambu-Goto string, one may obtain static 
%%strings in a spacetime. 

One may expect 
that if we assume ``less'' homogeneity, 
the equation still remains tractable and 
the solutions have enough variety to include 
nontrivial configurations and dynamics of physical interest. 
The {\co} objects give such a class, 
which helps us to understand the basic properties of the extended 
objects and serves as a base camp to explore their general
dynamics.  
For a string, stationarity is a special case of the {\conohyphen}
condition. Some stationary configurations of the 
Nambu-Goto strings are 
obtained in the Schwarzchild spacetime~\cite{FSZH}. 
%%\fbox{Older examples}
Even in the Minkowski space, 
many nontrivial {\co} solutions of the string 
were recently found~\cite{ogawa,IshKoz05PRD}. 
A {\co} object is defined, roughly speaking, as the one whose 
world sheet is homogeneous except in one direction. 
Then any covariant PDE governing such an object reduces
to an ordinary differential equation (ODE), which can easily be solved
analytically, or at least, numerically. 
A solution represents the dynamics of a spatially homogeneous object, 
or the nontrivial configuration of a stationary object, depending on 
the homogeneous ``direction'' is spacelike or timelike. 
The case of null homogeneous ``direction'' should also give new 
intriguing models. 

%% Let $(\M,G)$ be a {\it geometry}, 
%% i.e.,  $\M$ is a manifold and $G$ is a Lie group acting transitively
%% on $\M$. 
%% Let $\sfc$ be a submanifold of $\M$. 
%% Let $H$ be a subgroup of $G$ which preserves $\sfc$. 
%% We call that $\sfc$ is {\em homogeneously
%%   embedded}\/ to $\M$ if  $H$ acts transitively on $\sfc$. 
%% We define, in general, the {\em homogeneity}\/ of $\sfc$ by 
%% \begin{align}
%%   h=\max\{\dim H(p); p\in \sfc\}, 
%% \end{align}
%% and the {\em cohomegeneity}\/ of $\sfc$ by $\dim S-h$. 
%% A homogeneously embedded submanifold has 
%% cohomogeneity zero. 

%% Below we will focus on the case where $\M$ is a pseudo-Riemannian 
%% manifold and $G:=\Isom_0M$, 
%% %%These do not change if we restrict our attention to 
%% the identity
%% component of the isometry group of $\M$.  

In this paper, we treat strings 
in the five-dimensional anti-de Sitter space $\AdS^5$.
The choice of the spacetime 
is to meet the recent interest in higher-dimensional
cosmology, including the brane-world universe model, 
and in string theory, though the method developed here is applicable
to any background spacetime. 
%% Furthermore, among extended objects, we focus on two-dimensional
%% ones, namely,  strings. 
%% The solutions can be used to investigate dynamics
%% of the  strings in the higher-dimensional cosmology. 
A particular example which has recently been attracting much attention is 
the  string in a spacetime with large extra 
dimensions, which are suggested e.g. by the brane-world model. 
A detailed investigation~\cite{Jackson:2004zg} suggests that 
the reconnection probability for this type of strings is 
significantly suppressed. Then, contrary to what had usually been believed, 
the strings in the Universe can stay long enough 
to be considered stationary. 
Therefore classifying {\co} strings and solving dynamics thereof are 
important for examining the roles of the string in cosmology. 
%%, because stationarity implies 
%%{\co} property for  strings. 
We first give the classification of all {\co}  strings 
which is valid for any covariant equation of motion. 
Then, in the case of Nambu-Goto strings, 
we give a general method for solving 
the trajectory. 
The method can be easily applied to the cases of other equations of motion. 
We demonstrate the procedure and give explicit
solutions in some particular cases. 

In the classification, we make use of the 
local isomorphism between $\SO(4,2)$ and $\SU(2,2)$ in an essential
way. 
%% as well as in solving the equation of motion. 
The latter group is easier to treat 
%%for the reasons including: 
%%the action on the space $\E^{4,2}$ can be written consicely with a natural
%%complex coordinates, 
because the dimensionality of the matrix is lower 
and because 
the Jordan decomposition of complex matrices is simpler than that
of real ones. 
Therefore, though a similar classification of Killing fields 
is found in literature in the context of constructing quotient
spaces of the anti-de Sitter space~\cite{HolPel97CQG}, 
we present an alternative proof based on the classification of 
$H$-{\asa} matrices in the Appendix. 
%%(Sec.~\ref{HSA}) 
%% and 
%% the classification %(Sec.~\ref{classification}) 
%% of the pair $(A,H)$ 
%% of a $H$-{\sa} matrix $A$ and a Hermitian matrix $H$ 
%% (Sec.~\ref{classification}) 
%% turn out to be essential. 

In Sec.~\ref{general}, we give a method 
for the classification of all {\co} strings in general, and 
a method for solving the equations of motions for Nambu-Goto strings. 
The latter can be easily applied to other equations of motion. 
In Sec.~\ref{loc iso}, The useful relation of the isometry group $\Go$
and $\G$ is briefly explained. 
We give the classification of the {\co} strings in the anti-de Sitter
space in Sec.~\ref{classification}. 
In Sec.~\ref{hopf}, we demonstrate 
the method presented in
Sec.~\ref{general} by an example. 
There we solve the Nambu-Goto equation and examine the geometry of
its world sheet. 
Sec.~\ref{conc} is devoted for conclusion.

In this paper, 
a spacetime $(\M,g)$ is a manifold $\M$ endowed with a
Lorentzian metric $g$. 
We denote  by $G$ the identity component of the isometry group of
$(\M,g)$, and by $\alg g$ its Lie algebra. 
We use the unit such that the speed of light and Newton's constant
are one.

\begin{figure}[htbp]
  \centering
          \includegraphics[width=.6\linewidth]
          {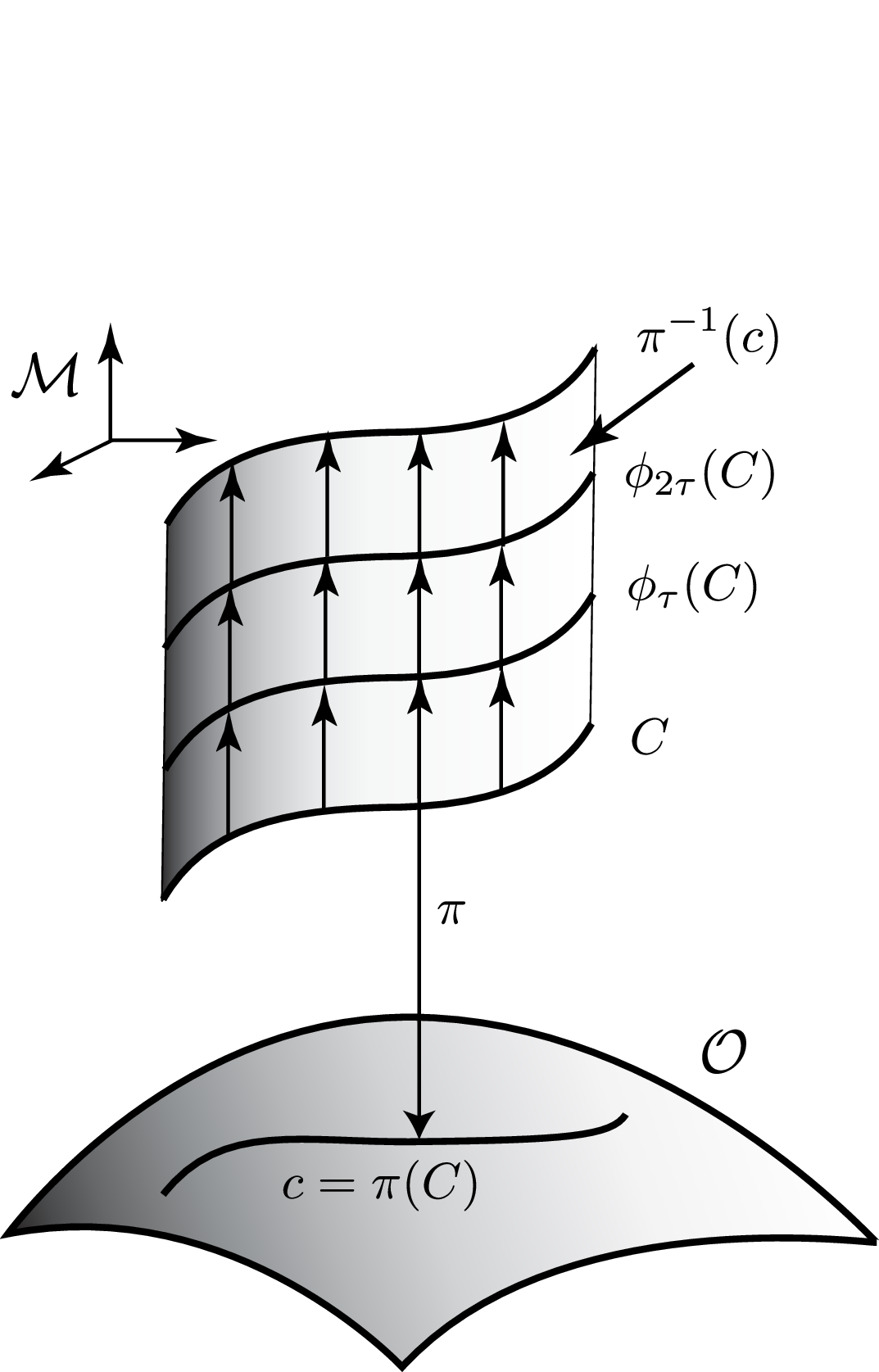} 
          \caption{To solve a trajectory of the {\co}  string is
          to find a curve $C$ in $\M$ which projects to a
          geodesic $c$ on $\Or$.}
  \label{fig-bundle}
\end{figure}

\section{General treatment of {\co}  strings}
\label{general}
In this section, we develop a general method for classifying 
{\co} objects and solving their dynamics in an arbitrary spacetime
$(\M,g)$. 
Let us start with the definition of the {\co} objects. 
We say that a $m$-dimensional hypersurface $\sfc$ in $\M$ 
is of {\it \conohyphen}\/ if it is foliated by 
$(m-1)$-dimensional submanifolds $\sfc_\sigma$ labeled by a real
number $\sigma$ and there is a subgroup $K$ of
$G$ which preserves the foliation and acts transitively on $\sfc_\sigma$. 
In particular, the hypersurfaces $\sfc_\sigma$'s are 
embedded homogeneously in $\M$. 
A {\co} object has a world sheet which is a {\co} hypersurface. 
In this paper, we focus on the case that the extended objects are
 strings, so that  $m=2$, and 
$K$ is a one-parameter group $(\phi_\tau)_{\tau\in\R}$ of isometries. 

First, let us consider how to classify the {\co}  strings. 
Given a one-dimensional subgroup $K\subset G$ and a point $p\in M$, 
the equations of motion determines 
a unique world sheet of a {\co} object. 
The dynamics of the two  strings can be considered the same if there
is an isometry sending one of their trajectories, $\sfc$, 
to the other, $\sfc'$. 
In this paper, we identify the two dynamics if we can do so gradually,
namely, if there is a one-parameter group of isometries
$(\phi'_\lambda)_{\lambda\in[0,1]}$ 
such that $\phi'_0$ is the identity and $\phi'_1(\sfc)=\sfc'$. 
We therefore 
classify the {\co}  strings up to isometry connected to the
identity. 
In terms of Killing vector fields, 
it is to classify the Killing vector field $\xi$ generating $K$ 
up to scalar multiplication and up to isometry. 
Namely, $\xi$  and $a\phi_*\xi$ are equivalent if there exists 
$\phi\in G$ and $a\ne0$. 
To put it more algebraically, the task is to find 
%% the equivalence class $\alg k/\Ad_G$, 
%% where $\alg k$ is a one-dimensional Lie subalgebra of $\alg g$ 
%% and $\Ad_g(X):=gXg^{-1}$,
%% $X\in\alg g$. 
%% It is in turn equivalent to find 
$\alg g/\Ad_G$ 
up to scalar multiplication. 

Second, let us give a formalism to solve the dynamics and the configuration of
the {\co}  strings. 
We assume that the string is described by the Nambu-Goto action 
\begin{align*}
  S=\int_S\sqrt{-g_{ab}dx^adx^b}. 
\end{align*}
%%a one-parameter
%%family $(\phi_\tau)_{\tau\in\R}$ of isometries, 
The orbit space of the string with the symmetry group $K$ 
is defined by $\Or:=\M/K$, i.e., 
by identifying all the points on each Killing orbit in $\M$. 
The submanifolds $\sfc_\sigma$ mentioned above are 
the preimages $\pi^{-1}(x)$ of a point $x\in\Or$. 
One can endow $\Or$ with a metric $h$ so that the projection 
$\pi: (\M,g)\to (\Or,h)$ is an orthogonal projection, or more precisely, 
 %%(pseudo-)
a Riemannian submersion. The metric $h$ is given by 
\begin{align}
  h_{ab}:=g_{ab}-\xi_a\xi_b/f, 
  \label{hab}
\end{align}
where $f:=\xi^a\xi_a$. 
This metric has the Euclidean signature if 
the Killing vector $\xi$ is timelike, i.e., if $f<0$, 
and the Lorentzian signature if $\xi$ is spacelike, i.e., if $f>0$. 
Carrying out the integration along $\xi$ in the Nambu-Goto action, 
one obtains 
\begin{align}
  S=\int_c \sqrt{-f h_{ab}dx^adx^b}, 
  \label{line_action}
\end{align}
where $c$ is a curve on ${\Or}$. 
Thus the problem of the  string reduces to finding geodesics on 
the orbit space $\Or$ with the metric $-fh$. 
For convenience, we adopt a modified action 
\begin{align}
  S=\int_c 
  d\sigma\paren{-\frac1{\alpha}f h_{ab}\dot x^a\dot x^b+{\alpha}}, 
  \label{line_action_2}
\end{align}
where an overdot denotes the differentiation by $\sigma$. 
The action \Ref{line_action_2} derives the same geodesic equations as
\Ref{line_action} and retains 
the invariance under reparametrization of $\sigma$. 
The function ${\alpha}$ is the norm of the tangent vector. 

The two-dimensional 
world sheet of the  string is the preimage $\pi^{-1}(c)$ of
the geodesic $c$ on $(\Or,-fh)$. 
However, it is sometimes more convenient to find a {\em lift}\/ curve
${C}$ on $\M$ whose 
projection $\pi({C})$ is a geodesic on $(\Or,-fh)$ 
than to find a geodesic on $(\Or,-fh)$ 
(Fig.~\ref{fig-bundle}). 
The Hopf string in Sec.~\ref{hopf} is such an example.
%% where one can choose ${C}$ as a geodesic on $(\M,g)$. 
In the case, the trajectory of the string is given by 
\begin{align}
  \sfc
  &=\pi^{-1}(\pi({C})) 
  \nn
  &= 
  \{\phi_\tau({C}(\sigma));\, (\tau,\sigma)\in \R^2\}. 
\label{eq-general-trajectory}
\end{align}
%%where $U\subset\R^2$. 
Note that the last expression in 
\Ref{eq-general-trajectory} 
depends on the objects in $\M$ only. 
Thus the trajectory $\sfc$ can be viewed as a foliation by 
mutually isometric curves $\phi_\tau\circ{C}$ 
labeled by $\tau$. 

%%\fbox{homogeneity-preserving isometry}
After one obtains the solutions of the equation of motion, 
one may want to classify their trajectories up to isometry. 
This can be done by identifying $C$ (or $\sfc$) which are related by {\em
  homogeneity-preserving isometries}. We say that an isometry 
$\Phi$ is homogeneity-preserving if it preserves the action of $K$,
i.e., if it satisfies 
\begin{align}
  \Phi\circ K\circ\Phi^{-1}=K. 
\end{align}
The homogeneity-preserving isometries form a group. 
In algebraic terms, the group is the normalizer of $K$ in the group $G$ of
isometries on $\M$, which is denoted by $N_G(K)$. 
Its Lie algebra is the idealizer of ${\alg k}$ in ${\alg g}$ which is denoted by
$N_{\alg g}(\alg k)$. 
%%The group $N_G(K)$ acts on $(O)$ 

%% The {\em centralizer} $Z_G(K)$, 
%% which consists of the elements $\Phi$ 
%% of $G$ commuting with all elements of $K$, 
%% is a subgroup of $N_G(K)$. 
We note that in the special case that $\Phi$ commutes with the action of $K$, 
{\ie} when $\Phi$ is in the {\em centralizer}\/ $Z_G(K)$ of $K$ in $G$, 
the squared norm of $\xi$ must be invariant under $\Phi$. 
This can be seen from 
$\Phi_*f
=\Phi_* (g_{ab}\xi^a\xi^b)
=(\Phi_*g_{ab})\xi^a\xi^b
+g_{ab}(\Phi_*\xi^a)\xi^b
+g_{ab}\xi^a(\Phi_*\xi^b)
=f
$, 
where we have used $\Phi_*g_{ab}=g_{ab}$ and 
$\Phi_*\xi^a=\xi^a$.

%% We note that the normalizer $N_G(K)$ equals the {\em centralizer} 
%% $Z_G(K)$ which consists of the elements $\Phi$ 
%% of $G$ which commutes with all elements of $K$, 
%% if $K$ is one-dimensional and $f$ is nonzero. 
%% This follows from 
%% $\Phi_*f
%% =\Phi_* (g_{ab}\xi^a\xi^b)
%% =(\Phi_*g_{ab})\xi^a\xi^b
%% +g_{ab}(\Phi_*\xi^a)\xi^b
%% +g_{ab}\xi^a(\Phi_*\xi^b)
%% $ .....

The whole procedure of solving the dynamics is explicitly carried out
for an example in Sec.~\ref{hopf}.

%%\section{$\Goo$ and $\G$}
\section{$\AdS^5$ and its isometry group}
\label{loc iso}
Hereafter in this paper, we assume that the spacetime $(\M,g)$ is 
the five-dimensional anti-de Sitter space $\AdS^5$, or its universal
cover $\wt{\AdS^5}$. The former space 
has closed timelike curves which in 
the latter space are ``opened up'' to infinite nonclosed curves. 
The latter is usually more suitable when we discuss cosmology, 
but we will not distinguish them strictly in the following. 

The space $\AdS^5$ is the most easily expressed as a pseudo-sphere 
%%$-\s^2-t^2+w^2+x^2+y^2+z^2=-l^2$ 
\begin{align}
  \wb {\psi} {\psi}=-1
  \label{eq-AdS}
\end{align}
in the pseudo-Euclidean space $\E^{4,2}$ whose metric is 
$dS^2=l^2d\wb {\psi}d{\psi}$, 
where we have used complex coordinates 
${\psi}:=({\psi}^0,{\psi}^1,{\psi}^2)^T\in\C^3$, 
and have defined 
$\wb {\psi}:={\psi}^\dagger\zeta$ and 
$\zeta:=\diag[-1,1,1]$. 
%%\fbox{what are the connected components of $\Go$?}

The isometry group of $\AdS^5$ is $\Goo$ 
acting on $(s,t,x,y,z,w)^T\in\R^6$, 
where ${\psi}^0:=s+it$, 
${\psi}^1:=x+iy$, and
${\psi}^2:=z+iw$. 
In the classification of the  strings, however, 
we take advantage of the
isomorphism $\Go\simeq\G/\{\pm1\}$ and work with $\G$. 
%% so that 
%% we shall classify $\go/\Ad_{\Go}$. 
%% To do so, 
%% we take advantage of the isomorphism 
%% $\Go\simeq \G/\{\pm1\}$. 
%% This isomorphism implies 
%% in particular 
%% a Lie algebra isomporphism 
%% $\go\simeq\g$ 
%% and a Lie group isomorphism
%% $\Ad_{\Go}\simeq \Ad_{\G}$. 
%% Thus we have an isomorphism 
%% $\go/\Ad_{\Go}\simeq\g/\Ad_{\G}$, 
%% %%We shall classify $\g/\Ad_{\G}$ in the sequel. 
%% which will be resolved in Sec.~\ref{classification}. 
%%
%%Let us give a concrete expression of the $\Go$ transformation by $\G$. 
Let $V$ be the vector space whose elements are complex antisymmetric
matrices of the form 
\begin{align}
  p
  &=
  \begin{bmatrix}
    0&\conj{({\psi}^0)}& \conj{(\psi^1)} & -{\psi}^2\\
    -\conj{({\psi}^0)}&0&-\conj{(\psi^2)} & -{\psi}^1    \\
    -\conj{(\psi^1)} & \conj{(\psi^2)}&0&-{\psi}^0\\
    {\psi}^2 & {\psi}^1&{\psi}^0&0
  \end{bmatrix}
  \nn
  &=s i\pa z\ot\pa y+t 1\ot\pa y
  +x i\pa y\ot\pa z
  \nn
  &\qquad
  +y\pa y\ot 1
  -zi\pa y\ot\pa x+w \pa x\ot \pa y, 
  \label{eq-SU22rep-2}
\end{align}
where $\pa x$, $\pa y$ and $\pa z$ are the Pauli matrices and $1$ is the 
$2\times2$ identity matrix. 
The action of an element of $\Go$ on $E^{4,2}$ 
corresponds to the action of $U\in\G$ 
on $V$ in the following way~\cite[p106]{Yok90}: 
\begin{align}
  &p\mapsto UpU^T.
  \label{eq-SU22rep-1}  
\end{align}

The Lie algebra $\g$ of $\G$ consists of the matrices $X=$ satisfying 
$X\eta+\eta \da X=0$, where ${\eta}:=\diag[1,1,-1,-1]$. 
The explicit form is 
\begin{align}
  X=\mtx{\beta}{\gamma}{\da \gamma}{\delta}, 
\end{align} 
where $\gamma$ is a $2\times2$ complex matrix, and 
$\beta$ and $\delta$ are $2\times2$ anti-Hermitian matrices. 
The infinitesimal transformation for \Ref{eq-SU22rep-1} is given by 
the action of $X\in\g$ as
\begin{align}
  p\mapsto Xp+pX^T=\{X_S,p\}+[X_A,p], 
\end{align}
where 
$X_S:=(X+X^T)/2$ and $X_A:=(X-X^T)/2$ are the symmetric and
antisimmetric parts, respectively, of $X$. 
%% Let us express the correspondence in terms of a basis of 
%% $\g$. 
%% A convenient basis for $\g$ is 
%% $\{1/i,\pa x,\pa y,\pa z/i\}
%% \otimes
%% \{1,\pa x,\pa y,\pa z\}$ 
%% with $(1/i)\otimes 1$ being omitted 
%% [Including $(1/i)\otimes 1$ would yield a basis for $\alg u(2,2)$]. 
%% Then the infinitesimal transformation 
%% are expressed by a commutator 
%% $[X,\bullet]$ for 
%% $X=
%%   \{1/i,\pa x,\pa z/i\}
%%   \otimes
%%   \{1,\pa x,\pa z\}$ 
%% and for $X=   \pa y\otimes\pa y$, 
%% and by an anticommutator 
%% $\{X,\bullet\}$ for
%% $X=
%%   \pa y
%%   \otimes
%%   \{1,\pa x,\pa z\}$ 
%% and for $X=    \{1/i,\pa x,\pa z/i\}\otimes\pa y$. 
%% The corresponding $\Go$ infinitesimal transformations 
%% (Killing vectors on $\AdS^5$) 
%% are given in Table~\ref{tab-cor}, 
%%with $X$ being $(e_1\ot e_2)/2$. 
The correspondence between the $\g$ and $\go$ infinitesimal
transformations 
are given in Table~\ref{tab-cor}, 
where $X=(e_1\ot e_2)/2$. 
%% In the table, $xy$ denotes the rotation in the $xy$ plane, 
%% $st$ denotes the rotation in the $st$ plane, 
%% $tx$ denotes the $t$-boost in the $x$ direction, 
%% $-sw$ denotes the $s$-boost in the $-w$ direction, 
%% and so forth. 
%% \begin{table}[htbp]
%%   \centering
%%   \begin{ruledtabular}
%%     \begin{tabular}{c|cccc}
%%     \backslashbox{$e_1$}{$e_2$}  & 1& $\pa x$ &$\pa y$&$\pa z$\\ \hline
%%     $1/i$ & & $yz$ & $zx$ & $xy$\\
%%     $\pa x$ & $tw$ & $sx$ & $sy$ & $sz$ \\
%%     $\pa y$ & $-sw$ & $tx$ & $ty$ & $tz$ \\
%%     $\pa z/i$ & $st$ & $wx$ & $wy$ & $wz$ 
%%   \end{tabular}
%%   \end{ruledtabular}
%%   \caption{Correspondence between the $\g$ and $\go$ transformations.}
%%   \label{tab-cor}
%% \end{table}
In the table, $J_{xy}$ denotes the rotation in the $xy$ plane, 
$L$ denotes the rotation in the $st$ plane, 
$K_x$ denotes the $t$-boost in the $x$ direction, 
$\wt K_w$ denotes the $s$-boost in the $w$ direction, 
etc. 
\begin{table}[t]
  \centering
  \begin{ruledtabular}
    \begin{tabular}{c|cccc}
    \backslashbox{$e_1$}{$e_2$}  & 1& $\pa x$ &$\pa y$&$\pa z$\\ \hline
    $1/i$ & & $J_{yz}$ & $J_{zx}$ & $J_{xy}$\\
    $\pa x$ & $K_w$ & $\wt K_x$ & $\wt K_y$ & $\wt K_z$ \\
    $\pa y$ & $-\wt K_w$ & $K_x$ & $K_y$ & $K_z$ \\
    $\pa z/i$ & $L$ & $J_{wx}$ & $J_{wy}$ & $J_{wz}$ 
  \end{tabular}
  \end{ruledtabular}
  \caption{Correspondence between the $\g$ and $\go$ transformations.}
  \label{tab-cor}
\end{table}

%%\section{$H$-{\sa} matrices}
%%\label{HSA}
\section{The classification}
\label{classification}
In this section, 
we obtain the classification of the {\co}  strings in
$\AdS^5$. 
As discussed in Sec.~\ref{general}, 
%%above, 
the classification is to 
find $\alg g/\Ad_G$ up to scalar multiplication, where $G=\Go$. 
Because $\Go$ is isomorphic to $\G/\{\pm1\}$ as is seen in 
Sec.~\ref{loc iso}, 
$\go/\Ad_\Go$ is isomorphic to $\g/\Ad_\G$. 
Thus the classification is to find $\g/\Ad_\G$ up to scalar
multiplication. 
However, the equivalence classes $\g/\Ad_\G$ is known as in the Lemma
below, so that we can easily classify the 
{\co}  strings by further identifying the
equivalence classes by scalar multiplications. 

We begin with introducing some terms which is necessary to state the
Lemma. 
Let $H$ be an invertible Hermitian matrix. 
The {\it $H$-adjoint}\/ of a square matrix $A$ is defined 
by $\pad A:=H^{-1}\da AH$. 
A matrix $A$ is called {\it $H$-{\sa}}\/ when $\pad A=A$, 
{\it $H$-{\asa}}\/ when $\pad A=-A$, 
and {\it $H$-unitary}\/ when $A\pad A=\pad A A=1$. 
We say that matrices $A$ and $B$ are {\it $H$-unitarily similar} 
and write $A\simh B$ if 
there exists an $H$-unitary matrix $W$ satisfying $B=WAW^{-1}$. 
In these terms, 
$\G$ is the group of
unimodular ${\eta}$-unitary matrices 
%%with ${\eta}=\diag[1,1,-1,-1]$, 
and $\g$ is the Lie algebra of 
traceless ${\eta}$-{\asa} matrices.  
Thus, from the discussion in Sec.~\ref{general}, 
our task of classifying {\co}  strings is to 
classify the elements of $\g$
up to equivalence relation $\sime$ and 
up to scalar multiplication. 

%%\section{The classification}
%%\label{classification}
%%We shall now find out the equivalence classes of $\g/\Ad_{\G}=\g/\sime$. 
%%%%which is an equivalence class of $\g$ by 
%%%%the equivalence relation $\sim$, where $X\sim X'$ 
%%%%if there is $W\in\G$ such that $X'=WXW^{-1}$. 
%% {\ie}, find a canonical form for elements of $\g$ 
%% up to similarity transformations of $\G$. 
Let us introduce another equivalence relation closely related to the
one above. 
Let $(A,H)$ be a pair of a complex matrix and an invertible Hermitian
matrix $H$.  
The pairs $(A,H)$ and $(A',H')$ are said {\em unitarily
similar}\/ if there is a complex matrix $W$ such that
$%%\begin{align}
  A'=WAW^{-1}, 
  %%\quad
  H'=WHW^\dagger
%%\end{align}
$~\cite%[p31]
{GLR83}. 
This is an equivalence relation and will be denoted by 
$(A,H)\simu(A',H')$. 
Note that $A\sime A'$ is equivalent to 
$(A,{\eta})\simu(A',{\eta})$. 
Let $A$ be an $H$-{\sa} matrix. Then 
if $\lambda$ is an eigenvalue of $A$, so is its complex conjugate 
$\lambda^*$. 
Let $J_0(\lambda)$ be the Jordan block with eigenvalue $\lambda$ 
and let
\begin{align}
J(\lambda):=
\begin{cases}
  J_0(\lambda), & \text{$\lambda$ is real}, \\
  %%J_0(\lambda)\oplus J_0(\lambda^*), & \text{$\lambda$ is
  %%non-real}. 
  \diag[J_0(\lambda), J_0(\lambda^*)], & \text{$\lambda$ is non-real}. 
\end{cases}
\end{align}
%% $J(\lambda)$ be a Jordan block with eigenvalue $\lambda$ if 
%% $\lambda$ is real, 
%% and a direct sum of two Jordan blocks 
%% of the same size with eigenvalues
%% $\lambda$ and $\lambda^*$ if $\lambda$ is non-real.
Now we can state the Lemma~\cite%[Theorem 3.3]
{GLR83}.

\begin{lemma*}
%%{\it   
If $A$ is $H$-{\sa}, then 
%%$(A,H)$ is unitarily similar to $(J,P)$ with 
$(A,H)\simu(J,P)$ with 
\begin{align}
%% &J=J(\lambda_1)\oplus\cdots\oplus J(\lambda_\alpha)\oplus
%% J(\lambda_{\alpha+1})\oplus\cdots\oplus J(\lambda_\beta); 
%% \\
%% &P=
%% {\epsilon}_1 P_1\oplus\cdots\oplus {\epsilon}_\alpha P_\alpha\oplus
%% P_{\alpha+1}\oplus\cdots\oplus P_\beta; 
%%
%% J&=\bigoplus_{j=1}^{\beta}J(\lambda_j); 
%% \quad
%%  P=\Bigparen{\bigoplus_{j=1}^{\alpha}{\epsilon}_j P_j}
%% \oplus\Bigparen{\bigoplus_{j=\alpha+1}^{\beta}P_j}; 
%%
  J&=\diag\bracket{
   J(\lambda_1),\ldots, J(\lambda_\beta) }, 
 \\
 P&=
 \diag\bracket{
   {\epsilon}_1 P_1,\ldots, {\epsilon}_\alpha P_\alpha,
 P_{\alpha+1},\ldots, P_\beta } , 
\nn
{\epsilon}_j&=\pm1, 
\quad
P_j={\tiny \begin{bmatrix}
    {0}&{}&1\\{}&{\lddots}&{}\\1&{}&{\large 0}
  \end{bmatrix} }
                                     \text{ (antidiagonal)}
\end{align}
where 
$\lambda_1,\ldots,\lambda_\alpha$ are the real eigenvalues of $A$, 
$\lambda_{\alpha+1},\lambda_{\alpha+1}^*,\ldots,\lambda_\beta,\lambda_\beta^*$ 
are the non-real eigenvalues of $A$, and 
the size of $P_j$ is the same as that of $J(\lambda_j)$. 
%%}
\end{lemma*}
%%The representatives $(J,P)$ of the equivalence classes, which 
%%are called the canonical forms, are known [GLR, Theorem 4.1]. 
For any $X\in\g$, there is a pair $(J,P)$ in the Lemma such that
$(X/i,\eta)\sim(J,P)$, because $X/i$ is ${\eta}$-{\sa}. 
We will denote the {\em type}\/ of $X$ 
%%by the dimensions of $J(\lambda_j)$ for the real and non-real
%%eigenvalues of $X/i$, namely, 
by 
\begin{align}
  \text{\rm Type}(X):=
  ({\epsilon}_1d_1,\ldots {\epsilon}_\alpha d_\alpha|d_{\alpha+1}/2,\ldots,d_{\beta}/2), 
\end{align}
where $d_j:=\dim J(\lambda_j)$. [If there is either no real
($\alpha=0$) 
or no
non-real ($\alpha=4$) 
eigenvalues, we put a 0 in the corresnponding slot.]
We combine all the types with the same $d_j$ and call it 
the (major) type 
$[d_1,\ldots
d_\alpha|d_{\alpha+1}/2,\ldots,d_{\beta}/2]$, 
and we call $({\epsilon}_1,\cdots,{\epsilon}_\alpha)$ the minor type. 
In the Theorem below, 
$J_{xy}$ denotes spatial rotations in the $xy$ plane, 
$K_z$ denotes the boost with respect to the time $t$ in the $z$
direction, 
$\wt K_w$ denotes the boost with respect to the time $s$ in the $w$
direction, 
$L$ denotes the rotation in the $st$ plane, 
etc. 
%%and so forth. 

%%\begin{widetext}
\begin{table}[t]
\centering
\begin{ruledtabular}
  \begin{tabular}{ccc}
    %%\hline
  Type & Killing vector field $\xi$ 
  %%& Remark
  \\ 
  \hline
  $(4|0)$ & 
  $\disp {K_x+\wt K_y+J_{xy}+L}
  +2(J_{yz}+ K_z)$
  \\ 
  %%$(3,1|0)_{\pm}$
  $(\pm3,\mp1|0)$
  & 
  $\disp
  K_x+\wt K_y+J_{yz}
  \mp J_{xw}
  + a(J_{xy}-L\pm J_{zw})$
  %%& same amount of independent null rotations plus ...
  \\
  %%$(2,2|0)_{++}$
  $(2,2|0)$
  & 
  $K_x+ L+a J_{yz}$ 
  %%& a null temporal rotation, or a null rotation
  \\
  %%$(2,2|0)_{+-}$ 
  $(2,-2|0)$
  & 
  $K_x+J_{xy}+a J_{zw}$
  %%& a null temporal rotation, or a null rotation
  \\
  $(2,1,1|0)$ 
  & 
  $K_x+\wt K_y+J_{xy}+L+a\, J_{zw}+b\,(J_{xy}- L)$
  \\
  $(1,1,1,1|0)$ 
  & 
  $a\, L+b\, J_{xy}+c\,J_{zw}$ ($a^2+b^2+c^2=1$)
  %%& a temporal rotation and two rotations
  \\
  $(2|1)$ & 
  $K_x+\wt K_y+L+J_{xy}
  +a J_{zw} 
  + b(K_y+\wt K_x)$
  \\
  $(1,1|1)$ & 
  $K_x+\wt K_y+a\, J_{zw}+b\,(L-J_{xy})$
  \\
  $(0|2)$ 
  & 
  $K_x+J_{xy}+a\,\wt K_z$ ($a\ne0$)
  %%& a null rotation and a boost
  \\
  $(0|1,1)$ & 
  $aK_x+b\,\wt K_y+c\,J_{zw} \quad (b\ne\pm a,\  a^2+b^2+c^2=1)$
  %%& a spatial rotation and two boosts
  \end{tabular}
\end{ruledtabular}
\caption{The classification of {\co} strings. 
The types of the generator of $\G$ and the corresponding
  Killing vector fields $\xi$ on $\AdS^5$.} 
\label{tab-res}
\end{table}
%%\end{widetext}

%%Table~\ref\ref{tab-res-org} in the Appendix. 
\begin{theorem*}
\label{th-main}
%% Any generator of one-dimensional Lie subalgebra of $\go$ is 
%% equivalent to one of $\xi$ 
%% in Table~\ref{tab-res} up to conjugation of $\Go$, 
%% where $a,b,c\in\R$ and 
%% the double-signs must be taken in the same order in each expression.
%%
%% Any one-dimensional homogeneously embedded submanifold of $\AdS^5$ 
%% is generated by one of the nine types of $\xi$ 
%% in Table~\ref{tab-res} up to isometry of $\AdS^5$, 
%% where $a,b,c$ are real numbers, 
%% the double-signs must be taken in the same order in each expression.
%%
Any one-dimensional connected Lie group of isometries of $\AdS^5$ 
is generated by one of the nine types of $\xi$ 
in Table~\ref{tab-res} up to isometry of $\AdS^5$ connected to the
identity, 
where $a,b,c$ are real numbers, and 
the double-signs must be taken in the same order in each expression.
\end{theorem*}

The proof is given in the Appendix. 
Note in Table~\ref{tab-res} that 
Type $(0|1,1)$ would become Type $(1,1,|1)$
(with $b=0$) if one set $b=\pm a$ 
and that Type $(0|2)$ would become Type $(2,-2|0)$ (with
$a=0$) if one set $a=0$.

\section{The Hopf string}
\label{hopf}
In this section, 
we choose a type from the classified strings in
the Theorem and find its trajectory. 
We assume that the string obeys the Nambu-Goto equation and 
apply the general procedure presented in
Sec.~\ref{general}. 
The example also shows that 
working with the lift curves as explained in Sec.~\ref{general} can
make the calculations and
geometric interpretation of the trajectory 
simple and transparent. 

We shall say that a {\em Hopf string}\/ is a {\co}  string 
which is homogeneous under 
the change of the overall phase
in the complex coordinates $\psi$ defined in Sec.~\ref{loc iso}: 
\begin{align}
  {\psi}\mapsto e^{i\tau}{\psi}, 
  \quad \tau\in\R.
  \label{eq-hopf-symmetry}
\end{align}
This isometry is the simultaneous rotations
in the $st$, $xy$, and $zw$ planes. 
The Killing vector field 
$\xi$ is proportional to $L+J_{xy}+J_{zw}$ and falls into Type
$(1,1,1,1|0)$ with the condition $a=b=c$. 
The Killing orbits are closed timelike curves in $\AdS^5$. 
In the universal cover $\wt{\AdS^5}$, 
they are not closed and 
the string solution represents a stationary  string. 

Let us find the configurations of the Hopf string 
%% according to the procedure in Sec.~\ref{general}, 
%% {\ie}, 
by solving the action principle \Ref{line_action_2} 
and finding the geodesics on $(\Or,-fh)$. 
We first see that the orbit space $(\Or,h)$ is a Riemannian manifold, 
since $\xi$ is timelike. 
Then, from the fact that $f=\xi^a\xi_a$ is a constant (which we set $-1$), 
we find that 
solving the geodesics on $(\Or,-fh)$ is nothing 
but solving geodesics on $(\Or,h)$. 
One could either introduce some coordinate system on $\Or$ 
to solve \Ref{line_action_2} directly 
or make an ansatz with some coordinate system on $\AdS^5$ 
to solve \Ref{line_action}. 
Both methods work well but would lead to somewhat complicated equations. 
In what follows, we would take the advantage 
of the symmetry, especially the complex structure, 
of $E^{4,2}$ and find the lift curves on the spacetime $\AdS^5$ 
which project to the geodesics on 
$(\Or,-fh)$, as was explained in Sec.~\ref{general}. 

%% Let us examine the geometry of $(\Or,h)$. 
%% It is more convenient to work with the coordinates ${\psi}$ 
%% than to introduce a coordinate system on $\Or$. 
The metric $h$ in \Ref{hab} for the Hopf string is 
the usual flat metric $d\wb\psi d\psi$ 
with the contribution from 
the phase change being subtracted. 
With the constraint \Ref{eq-AdS}, 
$h$ can be written as 
\begin{align}
  h=l^2d\wb {\psi}(1-P)d{\psi},  
\end{align}
where $P:=-{\psi}\wb {\psi}$ is the normal projection along
${\psi}$. 
This is the same as the Fubini-Study metric on a projective space $\C P^2$ 
except that we started with an
indefinite scalar product $\zeta=\diag[-1,1,1]$ in \Ref{eq-AdS} and in
$dS^2=l^2d\wb\psi d\psi$, 
while the usual Fubini-Study metric is defined by means of a 
positive definite scalar product. 
We shall also call 
$h$ as the Fubini-Study metric here and shall denote 
the Riemannian manifold $(\Or,h)$ by $\C P_-^2$. 
The fibration %%$U(1)\to \AdS^5\to \C P_-^2$ 
$\C P_-^2\simeq \AdS^5/U(1)$ 
is the generalization of the Hopf fibration
to the case of indefinite scalar product~\cite{fn-a}. 
Thus the problem of finding Nambu-Goto strings has reduced to solving 
geodesics on %%$(\Or,-fh)=(\Or,h)$. 
$\C P_-^2$. 

%%From the argument in Sec.~\ref{general}, 
Our action \Ref{line_action_2} for the Hopf
string becomes 
\begin{align}
  S=\int_C d\sigma\paren{
    \frac1{\alpha}\dot{\wb {\psi}}(1+{\psi}\wb {\psi})\dot{{\psi}}+{\alpha} 
  +\mu(1+\wb {\psi}{\psi})}, 
  \label{line_action_hopf}
\end{align}
where $\mu$ is a Lagrange multiplier. 
This is the action for geodesics on $\Or$ written in terms of the
coordinates $\psi$ %%in $\AdS$, or 
in $E^{4,2}$. 
The action \Ref{line_action_hopf} has a $U(1)$ gauge invariance 
${\psi}(\sigma)\mapsto e^{i\theta(\sigma)}{\psi}(\sigma)$ 
\cite{fn-c} 
which corresponds to the freedom in the choice of a lift. 
This gauge degree of freedom is used to simplify the calculation. 
In particular, we shall show that each geodesic on $\Or$ for the Hopf string 
can always be written in a proper gauge as the
projection of a {\em geodesic}\/ on $\AdS^5$. 

The Euler-Lagrange equations are 
the constraint \Ref{eq-AdS} and 
\begin{align}
  \dot{\wb {\psi}}(1+{\psi}\wb {\psi})\dot{{\psi}}={\alpha}^2, \qquad & 
  \label{eq-hopf-norm-Z}
  \\
  -\paren{\frac1{\alpha}(1+{\psi}\wb {\psi})\dot
    {\psi}}\parbox[c][4ex]{1em}{}^\bullet
  +\frac1\alpha\dot{\wb\psi}\psi\dot\psi
  +\mu {\psi}&=0. 
  \label{eq-hopf-ddZ}
\end{align}
Multiplying $\wb {\psi}$ on \Ref{eq-hopf-ddZ} from the left and using
the constraint \Ref{eq-AdS}, 
one obtains an equation which merely determines $\mu$. 
On the other hand, 
the time derivative of \Ref{eq-AdS} implies that $\wb {\psi}\dot {\psi}$
is pure imaginary. 
This value can be changed by the gauge transformation
${\psi}\mapsto e^{i\theta(\sigma)}{\psi}(\sigma)$. 
%%which leaves
%%invariant the projected point in $\Or$. 
We can always 
choose the gauge $\Re\wb {\psi}\dot {\psi}=0$ which under the constraint 
\Ref{eq-AdS} implies 
\begin{align}
  \wb {\psi}\dot {\psi}=0.   
  \label{eq-parallelity}
\end{align}
Geometrically, \Ref{eq-parallelity} means that 
the curve ${C}$ on $\M$ is {\em horizontal}, namely, 
it is orthogonal, with respect to $g$, 
to the fiber 
$\pi^{-1}(\pi\circ{C}(\sigma))$ at each point on ${C}$. 
%%In this gauge, 
Multiplying $1+{\psi}\wb {\psi}$ on \Ref{eq-hopf-ddZ} from the left, 
and using \Ref{eq-AdS} and \Ref{eq-parallelity}, 
one obtains 
the geodesic equation for the Fubini-Study metric,  
\begin{align}
    (1+{\psi}\wb {\psi})\paren{\frac{\dot
     {\psi}}{\alpha}}\parbox[c][4ex]{1em}{}^\bullet=0. 
  \label{eq-hopf-geodeq}
\end{align}

Choosing the parameter of the curve to be the proper length 
so that ${\alpha}\equiv1$, 
one can write 
\Ref{eq-hopf-geodeq} in a particularly simple form. 
Since \Ref{eq-hopf-norm-Z} and \Ref{eq-parallelity} imply 
$\wb {\psi}\ddot {\psi}=-\dot{\wb {\psi}}\dot {\psi}=-1$, 
\Ref{eq-hopf-geodeq} yields
\begin{align}
  \ddot {\psi}={\psi}. 
\end{align}
One can immediately solve the equation to obtain 
\begin{align}
  &{\psi}(\sigma)=A\cosh\sigma+B\sinh\sigma, 
  \label{eq-hopf-geod-lift}
  \\
  &\wb AA=-1, \quad
  \wb AB=0, \quad
  \wb BB=1, 
  \label{eq-hopf-geod-lift-cond}
\end{align}
where $A,B\in\C^3$. 
The projection $\pi\circ{C}$ of the curves 
${C}:\sigma\mapsto\psi(\sigma)$  
expressed by \Ref{eq-hopf-geod-lift} are
geodesics on $\Or$. 

Some remarks are in order. 
First, the geodesics on the four-dimensional manifold $\Or$ 
should contain seven independent real constants: 
the initial position and the direction of the 
initial velocity. 
One sees that $\pi\circ{C}$ actually contains 
seven independent real constants since we have 
twelve real constants, four constraints 
\Ref{eq-hopf-geod-lift-cond} and 
one redundancy, i.e., the phase of $\psi(0)$. 
%%which we shall fix by requiring that $\psi^0(0)$ be real. 
%% With the coordinates $z^j=\psi^j/\psi^0$ ($j=1,2$), we have
%% \begin{align}
%%   z^j=
%%   \frac{\psi^j(0)\cosh\sigma+\dot\psi^j(0)\sinh\sigma}
%%   {\psi^0(0)\cosh\sigma+\dot\psi^0(0)\sinh\sigma}
%% \end{align}
Second, the lift curve \Ref{eq-hopf-geod-lift} is a {\em horizontal 
  geodesic}\/ on %%$(\M,g)$. 
$\AdS^5$. 
A special feature of the Hopf string is that 
one can always choose a lift curve $C$---the horizontal lift in this case--- 
of a geodesic $c$ on the orbit space $(\Or,-fh)$ so that 
$C$ {\em is also a geodesic on $(\M,g)$}. %%the spacetime $(\M,g)$ 
%% [in terms of Fig.~\ref{fig-bundle}, 
%% a geodesic $c$ on $(\Or,-fh)$ 
%% is the projection of a geodesic $C$ on $(\M,g)$.]
%%This happens because $f$ is constant. 
%%which can be easily verified. 
%% that ${C}$ is a geodesic on $\AdS^5$. 
Third, 
a horizontal geodesic ${C}$ on $\AdS^5$ is the intersection of $\AdS^5$
and a two-dimensional plane through the origin in $E^{4,2}$, which
corresponds to the great 
circle in the case of positive definite metric. 
Thus the hyperbolic curve 
\Ref{eq-hopf-geod-lift} is unique up to isometry, 
for any choice of $A$ and $B$. 
Furthermore, ${C}$ is a Killing orbit on $\AdS^5$. 

Now the world sheet $\sfc$ of the Hopf string can be written down easily.  
From \Ref{eq-general-trajectory}, \Ref{eq-hopf-symmetry} and
\Ref{eq-hopf-geod-lift}, 
we have 
\begin{align}
  \psi(\tau,\sigma)=e^{i\tau}
  \paren{A\cosh\sigma+B\sinh\sigma}, 
  \label{eq-hopf-trajectory}
\end{align}
where $A$ and $B$ satisfy the condition \Ref{eq-hopf-geod-lift-cond}. 

\begin{figure}[t]
  \centering
          \includegraphics[width=.9\linewidth]
          {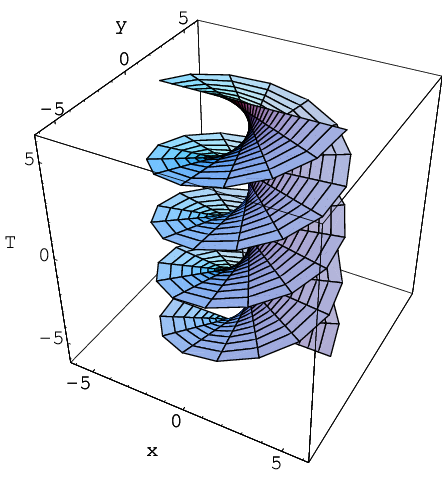} 
          \caption{The world sheet of the Hopf string in the case
          in the coordinates $(T,x,y)$. The other coordinates $z$ and $w$
          vanish. } 
  \label{fig-hopf}
\end{figure}

To describe geometry of the world sheet $\sfc$ 
in more detail, 
let us introduce a new time coordinate $T$ on $\AdS^5$ 
defined by 
\begin{align}
  T=\arg\psi^0=\arg(s+it). 
\end{align}
%% We take the hypersurface specified by $s=0$, $t>0$ in $\AdS^5$
%% (embedded in $E^{4,2}$) as 
%% the $\tau=0$ hypersurface. 
In $\wt{\AdS^5}$, $T$ runs from $-\infty$ to $\infty$. 
The $T=\const$ hypersurfaces embedded in $\wt{\AdS^5}$ are Cauchy surfaces. 
%% We define the coordinate system $(T,x,y,z,w)$ on $\wt{\AdS^5}$ 
%% such that $T=\const$ hypersurfaces coincide with $\tau=\const$
%% hypersurfaces. 
%% The vector field $\partial/\partial T$ is then orthogonal to
%% the $T=\const$ hypersurface at each point. 
The Killing field $\xi=d/d\tau$ drives 
the simultaneous rotations in the $xy$ and $zw$ planes 
while going up along the $T$ axis. 
Thus the world sheet of the Hopf string can be viewed pictorially as  
the surface swept by a boomerang
\Ref{eq-hopf-geod-lift} flying up %%(in the $T$ direction) 
while rotating (Fig.~\ref{fig-hopf}). 
%% (around the $T$ axis). 
%% Then the degree of freedom of the trajectory 
%% how the boomerang \Ref{eq-hopf-geod-lift} is put 
%% in the space $E^{4,2}$, 
%% because 
%% the curve \Ref{eq-hopf-geod-lift} is unique up to isometry. 

Let us reduce the degrees of freedom of 
$A$ and $B$ in \Ref{eq-hopf-geod-lift} by the 
homogeneity-preserving isometries and 
canonicalize them, as was explained in Sec.~\ref{general}. 
%%A concrete calculation show that all homogeneity-preserving isometries 
%%commute with the isometry $\phi_\tau$ corresponding to $\xi$. 
The Lie algebra $N_{\alg g}(\alg k)$ 
of the  {\HP} isometries 
is the vector space spanned by
\begin{align}
%%&L,\;J_{xy},\;J_{wz},\;
&\xi,\; L-J_{xy},\; L+J_{wz},\; 
J_{yz}+J_{wx},\;J_{zx}+J_{wy},\;\nn
&\wt K_z+K_w,\;K_z-\wt K_w,\;
\wt K_x+K_y,\;K_x-\wt K_y. 
\label{eq-gen}
\end{align}
In fact, all generators \Ref{eq-gen} commutes with $\xi$. 
%%commute with the isometry $\phi_\tau$ generated by $\xi$. 
%%
%% Now we show that the trajectory is in fact unique up to isometry. 
The isometries generated by \Ref{eq-gen} 
map the solution \Ref{eq-hopf-trajectory} 
to another isometric one. 
First, using the isometries generated by $L$, $J_{xy}$ and $J_{wz}$, 
one can make a general 
$A\in\C^3$ in \Ref{eq-hopf-geod-lift-cond} 
to be real, i.e., to have no $t$, $y$, $w$ components. 
Then, by using  
$\wt K_z+K_w$ and 
$\wt K_x+K_y$, 
one has $A=(1,0,0)^T$. 
Next, we canonicalize $B$ by the isometries which leaves this $A$ 
unchanged. 
By $\wb AB=0$, $B$ must have the form $B=(0,B^1,B^2)$. 
By using $J_{xy}$ and $J_{wz}$, one can make $B^1$ and $B^2$ real. 
Finally, by using $J_{zx}+J_{wy}$, one has 
%% \begin{align}
%%   A=(1,0,0)^T, \quad
$B=(0,1,0)^T$, 
%%\end{align}
where $\alpha\in\R$. 
As a result, the trajectory \Ref{eq-hopf-trajectory} can be written up to
isometry as
%% \begin{align}
%%   \psi(\tau,\sigma)=e^{i\tau}
%%   \Vtr 
%%   {\cosh\sigma+i\sinh\alpha\sinh\sigma}{\cosh\alpha\sinh\sigma}{0}. 
%% \end{align}
\begin{align}
  \begin{pmatrix}
    T\\x\\y\\z\\w    
  \end{pmatrix}
  =
  \begin{pmatrix}
    {\tau}\\
    {\sinh\sigma\cos\tau}\\
    {\sinh\sigma\sin\tau}\\
    0\\0
  \end{pmatrix}, 
  \label{eq-ws}
\end{align}
where we have used $T=\arg(s+it)$. 
In particular, the world sheet has no parameter and is unique. 
We can therefore say that the Hopf string has {\em rigidity}.

%% \begin{figure}[t]
%%   \centering
%%           \includegraphics[width=.9\linewidth]
%%           {fig-HopfString2-v2.eps} 
%%           \caption{The world sheet of the Hopf string in the case
%%           $\alpha=1$ in the $Txy$ space.} 
%%   \label{fig-hopf-2}
%% \end{figure}

Fig.~\ref{fig-hopf} shows the worls sheet of the Hopf string. 
This is a helicoid swept by a rotating rod passing through the
$T$  axis. This surface is periodic in $T$ direction with period 
$\pi$. 
The similar helical motion of an infinite curve in the Minkowski 
space has a cylinder outside of which the trajectory becomes
tachyonic (spacelike). 
In the anti-de Sitter case, however, the trajectory is always timelike
because the physical time passing with the unit difference in $T$
becomes large when the curve is far from the $T$ axis in
Fig.~\ref{fig-hopf}. 

Let us summarize some special features of 
the Hopf string. 
%%The symmetry has the following properties: 
(i) 
The Killing vector $\xi$ has a constant squared norm. 
%%In particular, $xi$ is always 
(ii)
The orbit space $(\Or,-fh)$ for Nambu-Goto Hopf string 
inherits the complex structure of $\E^{4,2}$, over which 
$\AdS^5$ admits a Hopf fibration. 
(iii)
The orbit space $(\Or,-fh)$ is homogeneous and is highly symmetric. 
(iv) The world sheet of the string is homogeneously embedded 
and is flat intrinsically. 
(v) 
The world sheet of the string is rigid, \ie, it is unique up to
isometry. 

%%We discuss the properties above. 
%%First, 
Among anti-de Sitter spaces, 
a Killing field satisfying 
(i) or (ii) exists 
only in the {\em odd-dimensional}\/ ones. 
%%It is trivial for (ii). 
%%As for (i), 
In the case of $\AdS^5$, 
the only Killing vector satisfying (i) 
is 
$L+J_{xy}\pm J_{zw}$ 
up to scaling and rotation of the spatial axes~\cite{fn-b}. 

The condition (i) is partially a reason for (ii) and (iii). 
%%(i) leads to $(\Or,-fh)=(\Or,h)$ and that 
In the case of the Hopf string, the homogeneity-preserving isometry 
group $N_G(K)$ equals the centralizer $Z_G(K)$. 
On the other hand, 
$Z_G(K)$ must preserve $f$ (Sec.~\ref{general}). 
Thus (i) in general suggests high symmetry of $(\Or,-fh)$. 
%%The orbit space $(\Or,-fh)$ of the Hopf string has 
In the case of Hopf string, the isometry group of the orbit space 
is an eight-dimensional group. 
In fact, 
The vector fields \Ref{eq-gen} except the first one $\xi$ 
form a closed Lie algebra and act on $(\Or,-fh)$ as Killing fields. 
%% All maps $\phi$ in the group preserve the fibers, 
%% i.e., the curves generated by $\xi$, 
%% hence correspond to isometries on $(\Or,h)$. 

As for (iv), 
one finds that the resulting 
world sheet \Ref{eq-ws} for the Hopf string is invariant
under the infinitesimal isometry $\wt K_x+K_{y}$ of $\AdS^5$. 
%%and thus can be considered as
%%a generalization of the solution in $\AdS^4$ to $\AdS^5$. 
Since $\xi$ and $\wt K_x+K_{y}$ commute, 
the world sheet $\sfc$ is acted by $\R^2$ 
and is homogeneous. 
This implies that $\sfc$ is flat intrinsically, namely, 
$\sfc$ is the two-dimensional Minkowski space embedded in $\AdS^5$. 
This can also be verified by a direct computation of the intrinsic metric. 

The high symmetry (iii) implies (v) for the Hopf string. 
Incidentally, stationary strings 
in $\AdS^4$ \cite{ads4} does not have rigidity. 
They would 
most naturally correspond in $\AdS^5$ to the cases $\xi\propto L+b J_{xy}$, 
which are in the same Type $(1,1,1,1|0)$ as the Hopf string 
but with different parameters. 
%%In fact, there is no Killing field of constant norm in $\AdS^4$. 

%% As is seen below, 
%% the constancy of $\xi^a\xi_a$ leads to high symmetry of the orbit space
%% $(\Or,-fh)$. 
%% This leads to the rigidity result for the Hopf string, 
%% {\ie}, the world sheet in $\AdS^5$ is {\em unique}\/ up to isometry and has no
%% parameter. 

These suggest that the Hopf string is similar to 
the string with simple time translation invariance in the Minkowski space. 
The Hopf string is the only solution in $\AdS^5$ which shares 
all of the properties (i), (iii), (iv) and (v) 
with the flat string in the Minkowski space.

\section{Conclusion} 
\label{conc} 
  The {\co} symmetry reduces the partial differential
  equation governing the dynamics of an extended object 
  in the spacetime $\M$ 
  to an ordinary differential equation. 
  With applications in higher-dimensional 
  cosmology in mind, 
  we have presented the procedure to classify 
  all {\co} strings and solve their trajectories with a given
  equation of motion. 
  The former is to classify the Killing vector fields up to isometry, 
  and the latter is to solve geodesics on the orbit space $(\Or,-fh)$
  which is the quotient space of $\M$ by the
  symmetry group $K$. 
  We have carried out the classification 
  in the case that the spacetime is the five-dimensional anti-de Sitter
  space, by an effective 
  use of the local isomorphism of $\SO(4,2)$ and $\SU(2,2)$ and 
  of the notion of $H$-similarity. 
  Assuming that the string obeys the Nambu-Goto equation, 
  we have solved the world sheet of one of the strings, 
  which we call the Hopf string, 
  in the classification. 
  The problem has reduced to find geodesics on the orbit space $(\Or,h)$. 
  By using a technique similar to the one used in quntum information
  theory and working on the lift curves in $\M$, 
  we have obtained a new solution which describes the 
  trajectories of the Hopf string.  
  They are timelike helicoid-like surfaces around the time axis 
  which is unique up to isometry of $\AdS^5$. 

We can say that the Hopf string is the simplest example of string in the 
anti-de Sitter 
space which corresponds to a straight static string in the Minkowski space. 
The Killing vector field defining the symmetry of the string is homogeneous
in the spacetime and has a constant norm. 
This greatly simplifies solving the geodesics on the orbit space
and the world sheet becomes homogeneous and rigid, 
as we have
seen in Sec.~\ref{hopf}. 
The simplicity of the Hopf strings suggests that they were common 
in the Universe and played significant roles, 
if the Universe is higher-dimensional or is a brane-world. 

We would like to remark that although we now have all types where 
the equations of motion reduce to ordinary differential equations 
this does not  in general imply solvability. 
The solvability problem is nontrivial and strongly related to the structure of
the orbit spaces. 
A systematic analysis will be presented in a future work. 

Finally, we would like to remark that 
the classification presented here will be the basis for that of
higher-dimensional {\co} objects.
The procedure is the following: 
(i) for each of the Killing vector field $\xi$ classified in
Table~\ref{tab-res}, 
enumerate how one can add new independent Killing vector fields  $\xi^{(1)}$, ...,
$\xi^{(n)}$ such that 
$\xi$, $\xi^{(1)}$ ..., $\xi^{(n)}$ form a closed Lie algebra $\alg
k'$; 
(ii) reduce the degrees of freedom of $\alg k'$ by using 
the isometries which preserve $\xi$, 
thus classifying the Lie algebras $\alg k'$; 
(iii) examine the orbits in the spacetime generated by $\alg k'$.

\section*{Acknowledgment}
The work is partially supported by 
Keio Gijuku Academic Development Funds (T. K.).

\appendix
\section*{Appendix: Proof of the Theorem}%~\ref{th-main}
\label{proof}
%%\begin{proof}%%[Sketch of proof]
Let $X$ be an ${\eta}$-{\asa} matrix $X$. 
%% Let us first show that ${\eta}$-unitarily similar to 
%% $X_0:=iW JW^{-1}$ with some $J$ in the Lemma and 
%% with an arbitrary complex matrix $W$ satisfying ${\eta}=W PW^\dagger$. 
The Lemma implies that 
$(X/i,{\eta})\simu(J,P)$ with some $(J,P)$. 
%% [This also implies that there exists at least one $W'$ satisfying 
%% ${\eta}=W' PW'{}^\dagger$.] 
On the other hand, if ${\eta}=W PW^\dagger$, 
the definition of unitary similarity implies
$(J,P)\simu(WJW^{-1},{\eta})$. 
Thus $(X/i,{\eta})\simu(WJW^{-1},{\eta})$ so that 
$X\sime iWJW^{-1}$. 
We therefore can carry out the classification by the following procedure: 
%%
%%\noindent 
(i) enumerate $(J,P)$ in the Lemma such that there exists $W$ satisfying 
${\eta}=WP\da W$, 
%%
%%\noindent 
(ii) construct $X_0=iWJW^{-1}$, 
%%
%%\noindent 
(iii) translate $X_0$ back to the Killing vector field $\xi$
in $\Go$ by Table~\ref{tab-cor}.  
%%Note that Step (i) restricts the minor type $({\epsilon}_1,...,{\epsilon}_\alpha)$ of
%%$X$ 

In some cases, however, the canonical pairs $(J,P)$ and $(J',P')$ 
correspond to $X_0$'s which generate an identical Lie group. 
This happens 
%% The degree of freedom is the magnitude
%% and the direction of the Killing field. 
%%The two pairs $(J,P)$ and $(J',P')$ generate an identical group 
when 
$(J',P')\simu(\alpha J,P)$ with 
a nonzero real number $\alpha$. 
Thus 
%% To find nontrivial identifications and 
%% clarify the relation between subtypes, 
it is important to know how a pair 
$(\alpha J_j(\la j),P_j)$ can be canonicalized. 
For $\alpha>0$, we simply have 
$(\alpha J_j(\la j),P_j)\simu(J_j(\alpha \la j),P_j)$, 
so that they generate an identical group. 
Thus we focus on $(- J_j(\la j),P_j)$ in the following. 
%%
%%\noindent (i) 
When $d_j$ is odd, we have 
\begin{align}
  \label{eq-dimP-odd}
  (-J(\la j),P_j)\simu(J(-\la j),P_j). 
\end{align}
This can be seen by applying a similarity transformation by 
$\diag[1,-1,1,\cdots]$. 
%%
%%\noindent (ii) 
When $d_j$ is even, we have 
\begin{align}
  \label{eq-dimP-even}
  (-J(\la j),P_j)\simu(J(-\la j),-P_j), 
\end{align}
which can be shown by applying 
a similarity transformation by $\diag[1,-1,1,\cdots]$, etc. 
%%
%%\noindent (iii) 
In the special case of $d_j=2$ and $\la j\in\C$, 
not only \Ref{eq-dimP-even} but also \Ref{eq-dimP-odd} holds 
because $-J(\la j)=J(-\la j)$. 

%%Another useful relation is that between $(J,P)$ and $(J,-P)$, 
The relation between $(J,P)$ and $(J,-P)$ is also important. 
Let us show that their corresponding Killing vector fields 
are related by a reflection $r:(t,x)\mapsto(-t,-x)$, 
which is a transformation in $SO(4,2)$ which is not connected to the
identity (hence is not used in the equivalence relation $\sime$).  
%%but not in $SO(4,2)_0$. 
When $(J,P)\simu(X_0,{\eta})$, we have 
$(J,-P)\simu(-X_0',{\eta})$ 
with $X_0':=-\Sigmaa X\Sigmaa^{-1}$ and $\Sigmaa:=\pa y\ot\pa x$, 
because $\Sigmaa {\eta}\da\Sigmaa=-{\eta}$. 
On the other hand, 
one can read off from \Ref{eq-SU22rep-2} that 
the transformation 
$p\mapsto-\Sigmaa p(\Sigmaa^{-1})^T=-\Sigmaa p\Sigmaa^T$ 
is a reflection along the $t$ and $x$ axes. 
Thus the Killing vector field $\xi$ corresponding to $X_0$ 
and the one $\xi'$ corresponding to $X_0'$ 
are related by $\xi'=r_*\xi$. 
%% If $X_0$ is invariant under $r$ (up to conjugacy of $\Go$ and up to
%% a scalar multiplication), $(J,P)$ and $(J,-P)$ result in 
%% generating the same one-dimensional subgroup of $\Go$. 

Let us find the relation of the minor types within each major type by
using the results above. We denote 
by an equal sign if two minor types are
related by $\eta$-unitary similarity which should be
considered identical, and 
by $\simsc$ if two minor types are
related by a scalar multiplication. 
%%\\
For Type $[4|0]$, it follows from \Ref{eq-dimP-even} 
that $(+)\simsc(-)$, 
which is invariant under $r$ (though the parameters change). 
%%\\
For Type $[3,1|0]$, 
there are two minor types $(+-)$ and $(-+)$ which are not related by
scalar multiplication but by the reflection $r$ (hence not equivalent
in the classification). 
%%\\
For Type $[2,2|0]$, 
it follows from \Ref{eq-dimP-even} that 
$(++)\simsc(--)$, 
which is invariant under $r$. 
by a simple reordering, we have $(+-)=(-+)$, 
which is invariant under $r$. 
%%\\
For Type $[2,1,1|0]$, 
by reordering, there are at most 
two  minor types $(++-)$ and $(--+)$. 
Furthermore, we have $(++-)\simsc(--+)$, 
by applying \Ref{eq-dimP-even} to all blocks. 
%% for the first block 
%% and by exchange of the second and the third blocks. 
It is invariant under $r$. 
%%\\
Type $[1,1,1,1|0]$ has only one minor type
(by reordering).
%%\\
For Type $[2|1]$, 
we have $(+)\simsc(-)$ 
by applying \Ref{eq-dimP-even} to the first block and 
\Ref{eq-dimP-odd} to the second block, 
yielding 
$(\diag[J_1,J_2],\diag[-P_1,P_2])\simu(-\diag[J_1',J_2'],\diag[P_1,P_2])$. 
%%\\
Type $[1,1|1]$ has a unique minor type 
(by reordering).
%%\\
Type $[0|2]$ and Type $[0|1,1]$ have a unique minor type. 

Let us demonstrate the concrete 
calculation for Type $[2|1]$ (the other types 
can be found in a similar manner). 
We have, because $J$ is traceless, 
$J=\diag\bracket{
  \mtx{a}10{a},
  -a+bi,-a-bi}$, 
where $a$ and $b$ are real numbers, and 
$P=\diag\bracket{
  \pm\mtx0110,
  \mtx0110}$. 
As discussed above, however, 
it suffices to consider the plus sign. 
Let us choose 
$W=S_{23}\cdot\diag[R(\pi/2),R(-\pi/2)]$ 
where $R(\theta)=
\begin{bmatrix}
  \cos\theta & -\sin\theta \\
  \sin\theta & \cos\theta
\end{bmatrix}
$
and
$S_{23}=
{\tiny
\begin{bmatrix}
  1&0&0&0\\
  0&0&1&0\\
  0&1&0&0\\
  0&0&0&1
\end{bmatrix} }. 
$ 
Then $X_0=iWJW^{-1}=
{\tiny
\begin{bmatrix}
  i(a+ 1/2) & 0 & i/2 & 0\\
  0 & -ia & 0 & b\\
  -i/2 & 0 & i(a- 1/2) & 0\\
  0 & b & 0 & -ia
\end{bmatrix}
}
=
-a\,(1/i)\ot\pa z+b\,\pa x\ot\frac{1-\pa z}2
- \frac{(\pa y+(\pa z/i))\ot(1+\pa z)}4
$. 
By Table~\ref{tab-cor}, we find that $X_0$ 
corresponds to the $\go$ transformation 
$
%%\begin{align}
  \xi=
  {-\wt K_w+K_z+L+J_{wz}}
  +a J_{xy} 
  + b(K_w-\wt K_z), 
%%\end{align}
$
where we have rescaled $\xi$ (by $-4$) 
and redefined $a$ and $b$. 
%%\end{proof}


\begin{thebibliography}{99}
\bibitem{fn-b} If one also admits the spatial reflection, 
  an isometry which is not connected to the identity, 
  one sees that $L+J_{xy}+J_{zw}$ is the only Killing vector of constant norm. 
  Alternatively, one can treat the case 
  with $L+J_{xy}-J_{zw}$ in the same manner as in the present section 
  by considering $(\psi^0,\psi^1,(\psi^2)^*)$ 
  instead of $(\psi^0,\psi^1,\psi^2)$.
\end{thebibliography}
\end{document}